\documentclass[reprint,aps,prb,longbibliography,letterpaper,final,twocolumn]{revtex4-2}
\usepackage{graphicx}
\usepackage{dcolumn}
\usepackage[colorlinks = true, linkcolor = blue, urlcolor  = blue, citecolor = blue, anchorcolor = blue]{hyperref}
\usepackage[english]{babel}
\usepackage{multirow}
\usepackage{dcolumn}

\begin{document}

\title{Thickness dependence of work function, ionization energy, and electron affinity of Mo and W dichalcogenides from DFT and GW calculations}
\author{Han-gyu Kim}
\author{Hyoung Joon Choi}
\email{h.j.choi@yonsei.ac.kr}
\address{Department of Physics, Yonsei University, Seoul 03722, Korea}

\begin{abstract}
Transition-metal dichalcogenides (TMDs) are promising for two-dimensional (2D) semiconducting devices and novel phenomena. 
For 2D applications, their work function, ionization energy, and electron affinity are required as a function of thickness, but research on this is yet to cover the full family of compounds.
Here, we present the work function, ionization energy, and electron affinity of few-layer and bulk 
$MX_2$ ($M=$ Mo, W and $X=$ S, Se, Te) in 2H phase obtained accurately 
by the density functional theory and GW calculations.
For each compound, we consider one-, two-, three-, four-layer, and bulk geometry. 
In GW calculations, accurate results are obtained by nonuniform $q$ sampling for two-dimensional geometry. 
From band energies including the GW self-energy correction, we estimate the work function, band gap, ionization energy, and electron affinity as functions of the number of layers. 
We compare our results with available theoretical and experimental reports, and we discuss types of band alignments in in-plane and out-of-plane junctions of these few-layer and bulk TMDs.
\end{abstract}

\maketitle

\section{Introduction}

Two-dimensional (2D) van der Waals materials are of great interest because of their variety of electronic properties and fabrication capability by mechanical exfoliation and stacking. 
While graphene is metallic with zero band gap, transition-metal dichalcogenides (TMDs) such as MoS$_2$ and WSe$_2$ are semiconducting with band gaps. 
For their electronic properties that depend sensitively on structures and chemical compositions, TMDs have been widely studied for device applications and novel phenomena~\cite{Mak2010,Splendiani2010,Britnell2013,Zhang2014,Xiao2012,
Xu2014,ZhangYJ2014,Mak2014, Kim2014,Sie2015,Ubrig2017,Onga2017,
Cao2012,Li2014,MacNeill2015,Aivazian2015,Srivastava2015,Tong2016,YeY2016, LeeJ2016}. 

One of the key properties of semiconducting materials is their electronic band-edge energies with respect to vacuum. 
They are related to the ionization energy, electron affinity, and work function, and also closely linked to band-edge alignment between different semiconductors, i.e., the band offset, required in designing semiconductor devices \cite{Pant2016,Bernardini1998,Walle1987,Wei1998,Kahn2016}. 
Previous theoretical studies of electronic band structures in TMDs have been based on the density functional theory (DFT) \cite{Kang2013,Gong2013,Gusakova2017,Zhu2011,Liu2013} and GW calculations \cite{Chernikov2014,Song2017,Ugeda2014,Naik2017,Naik2018,Bradley2015,Jiang2012,Liang2013}. 
Using DFT, band energies with respect to the vacuum level were studied for monolayers of $MX_2$ ($M =$ Mo, W and $X =$ S, Se, Te) \cite{Kang2013,Gong2013} and their bilayers, tetralayers, and bulks \cite{Kang2013}. 
The GW method, which introduces the self-energy due to electron-electron interaction \cite{Hedin1965,Hybertsen1985}, can describe band gaps in semiconductors more accurately \cite{Schilfgaarde2006}. 
The GW method was used to study band gaps, ionization energies, and electron affinities of bulk TMDs \cite{Jiang2012} and layers of MoS$_2$ \cite{Naik2017}, and band energies with respect to vacuum in monolayers of $MX_2$ ($M =$ Mo, W and $X =$ S, Se, Te) \cite{Liang2013}. 
For 2D applications of TMDs, their work function, ionization energy, and electron affinity are required as a function of thickness, but accurate study 
on this using the GW method is yet to cover the full family of compounds.

For a few layers of TMDs, GW calculation is computationally demanding. 
It is not only because a bigger unit cell is needed to host more atoms and wide-enough vacuum, but also because the poor electrical screening in 2D systems requires a much denser $q$-grid sampling for accurate convergence of the dielectric function \cite{Qiu2016}. 
Furthermore, band energies with respect to vacuum converge more slowly than the band gap, requiring more unoccupied bands in GW calculation \cite{Liang2013}. 
To study the thickness dependence of band structures of 2D TMDs with respect to the vacuum level, we relieve the requirement of a much denser $q$-grid sampling by using the nonuniform sampling of $q$-points, so-called, the nonuniform neck subsampling (NNS) method \cite{Jornada2017}, and we expedite the convergence of band energies with respect to the vacuum level by using the static remainder method \cite{Deslippe2013}.

In this paper, we study thickness dependence of band energies of TMDs in 2H phase with respect to vacuum, using DFT and GW calculations. 
We consider from monolayer (1L) to tetralayer (4L) and bulk of molybdenum and tungsten disulfide, diselenide, and ditelluride, i.e., $MX_2$ ($M =$ Mo, W and $X =$ S, Se, Te). 
In the case of GW calculations with 2D geometry, the NNS method and the static remainder method are used for accurate convergence of band energies. 
Including the GW self-energy correction to band energies, we obtain the ionization energy, electron affinity, work function, and band gap as functions of the layer thickness. 
From these properties, we also discuss possible band offsets in their in-plane and out-of-plane junctions.

\section{Methodology}

We performed DFT 
calculations using the QUANTUM ESPRESSO code~\cite{Giannozzi2009} with Perdew-Burke-Ernzerhof-type (PBE) generalized gradient approximation~\cite{Perdew1996} for the exchange-correlation energy. 
We use norm-conserving pseudopotentials for electron-ion interaction and plane waves with a kinetic energy cutoff of 125~Ry to expand electronic wavefunctions. 
For self-consistent calculations, we use a $15\times15\times3$ $k$-point sampling in the three-dimensional Brillouin zone (BZ) for bulk, and a $15\times15$ $k$-point sampling in the 2D BZ for few layers.

\begin{table}  
\caption{\label{tab1tmdc} 
Relaxed atomic structure of bulk TMDs using the PBE-D2 method. 
Experimental values from Refs. \cite{Wyckoff1963,James1963,Dawson1987}
are shown for comparison.}

\setlength{\tabcolsep}{1.1mm} 
\renewcommand{\arraystretch}{1.12}
\begin{tabular}{l r r r r r r}
\hline
&MoS$_2$&MoSe$_2$&MoTe$_2$&WS$_2$&WSe$_2$&WTe$_2$\\
\hline
PBE-D2& & & & & & \\
$a$ (\AA)&3.191&3.316&3.531&3.188&3.333&3.563 \\
$b$ (\AA)&12.417&13.032&14.018&12.160&12.799&13.802 \\
$u$ &0.625&0.622&0.620&0.621&0.620&0.619 \\
\hline
Experiment & & & & & & \\
$a$ (\AA)&3.160&3.289&3.518&3.180&3.290&3.600 \\
$b$ (\AA)&12.295&12.927&13.974&12.500&12.970&14.180 \\
$u$ &0.629&0.621&0.621&0.625&0.621&0.621 \\
\hline
\end{tabular}
\end{table} 

We performed GW calculations for quasiparticle band structures of bulk and few-layer TMDs using the BERKELEYGW code~\cite{Deslippe2012,Hybertsen1986,Rohlfing2000}. 
We used the one-shot GW method (G$_0$W$_0$) which uses DFT eigenvalues and wavefunctions as the starting point and calculates the self-energy once. 
Spin-orbit coupling is considered as a perturbation, after the self-energy correction is determined without spin-orbit coupling. 
We use the Godby-Needs generalized plasmon pole model \cite{Godby1989,Oschlies1995} for the frequency dependence of the inverse dielectric function. Our kinetic energy cutoff for the dielectric matrix is 35~Ry. 
We use a $6\times6\times2$ uniform $q$-point sampling for the bulk system. 
For few-layer TMDs, we use a $12\times12$ uniform $q$-point sampling in 2D BZ with an additional ten $q$-points determined by the NNS method~\cite{Jornada2017}, which is equivalent to a $2286\times2286$ uniform $q$-point sampling in the 2D BZ effectively. 
In 2D systems, unlike 3D systems, the inverse dielectric function has a sharp behavior near the ${\bf q} = 0$ point, requiring a very dense $q$-grid for converged quasiparticle energies. The NNS method improves the convergence near 
the ${\bf q} = 0$ point in 2D systems by including additional nonuniform 
$q$-points.
We include 3000 bands for all cases with the static remainder method~\cite{Deslippe2013}. 
Our calculation parameters show converged quasiparticle band energies within 0.1~eV.

We construct relaxed atomic structures of bulk and few-layer TMDs by minimizing their total energies within the PBE-D2 scheme~\cite{Grimme2006} which considers van der Waals interaction. 
For 2D TMDs, our supercells include large vacuum region of 25~\AA. In calculations of 2D TMDs, the vacuum level is obtained by averaging the electrostatic potential in the middle of the vacuum region. 
For the vacuum level of a bulk system, we simulate a 15-layer slab which is thick enough to show almost the same band gap as the bulk system with a difference smaller than 0.01~eV. 
At low temperature, we have the chemical potential right at the center of its band gap in an undoped semiconducting system. 
The work function, the ionization energy, and the electron affinity in DFT and GW calculations are given by absolute values of the chemical potential, 
conduction-band minimum (CBM), and valence-band maximum (VBM) obtained with respect to the vacuum level, respectively.

\begin{table} 
\caption{\label{tab2tmdc} Lattice parameters of few-layer and bulk TMDs. 
The height of the chalcogen atom from the metal plane is represented as $h_{M-X}$.}

\setlength{\tabcolsep}{1.5mm} 
\renewcommand{\arraystretch}{1.12}
\begin{tabular}{l c c c c}
\hline
Compound & Thickness & $a$ (\AA) & $b/2$ (\AA) & $h_{M-X}$ (\AA)  \\
\hline
\multirow{3}*{MoS$_2$}   
   &   1L           &   3.191    &            &   1.558    \\
   &   2L, 3L, 4L   &   3.191    &   6.210    &   1.558    \\
   &   bulk         &   3.191    &   6.209    &   1.558    \\
\hline
\multirow{2}*{MoSe$_2$}   
   &   1L                  &   3.316    &            &   1.666    \\
   &   2L, 3L, 4L, bulk    &   3.316    &   6.516    &   1.666    \\
\hline
\multirow{3}*{MoTe$_2$}   
   &   1L   &   3.530    &            &   1.818    \\
   &   2L, 3L, 4L   &   3.530    &   7.010    &   1.816    \\
   &   bulk   &   3.531    &   7.009    &   1.816    \\
\hline
\multirow{5}*{WS$_2$}   
   &   1L   &   3.189    &            &   1.568    \\
   &   2L   &   3.188    &   6.084    &   1.563    \\
   &   3L   &   3.189    &   6.086    &   1.564    \\
   &   4L   &   3.188    &   6.085    &   1.564    \\
   &   bulk   &   3.188    &   6.080    &   1.564    \\
\hline
\multirow{5}*{WSe$_2$}   
   &   1L   &   3.334    &            &   1.670    \\
   &   2L   &   3.334    &   6.401    &   1.665    \\
   &   3L   &   3.334    &   6.403    &   1.665    \\
   &   4L   &   3.333    &   6.401    &   1.665    \\
   &   bulk   &   3.333    &   6.400    &   1.665    \\
\hline
\multirow{5}*{WTe$_2$}   
   &   1L   &   3.566    &             &   1.808    \\
   &   2L   &   3.566    &   6.907    &   1.808    \\
   &   3L   &   3.566    &   6.909    &   1.806    \\
   &   4L   &   3.566    &   6.908    &   1.806    \\
   &   bulk   &   3.563    &   6.901    &   1.806    \\
\hline
\end{tabular}
\end{table} 

\section{Results and discussion}

We determined atomic structures of TMDs using the PBE-D2 method~\cite{Grimme2006,Giannozzi2009,Perdew1996} which considers van der Waals interaction. 
Table~\ref{tab1tmdc} shows relaxed and experimental atomic structures of bulk TMDs. 
We confirm that our DFT lattice parameters of bulk TMDs are consistent with experimental ones~\cite{Wyckoff1963,James1963,Dawson1987}
within 1\%-2\%. As shown in Table~\ref{tab1tmdc}, lattice parameters of bulk TMDs are weakly dependent on metal elements but strongly dependent on chalcogen elements. 
MoS$_2$ and WS$_2$ show very small differences less than, or close to, 1\% in $a$, $b$, and $u$. 
For heavier chalcogen elements, differences in $a$, $b$, and $u$ become a little bit larger, but still less than 2\%. 
This indicates that the atomic structure is almost determined by the chalcogen element only. 
We also relaxed atomic structures of TMD layers and obtained almost the same lattice constants, layer thicknesses, and interlayer distances as those of corresponding bulk structures, as compared in Table~\ref{tab2tmdc}.

\begin{figure} 
\includegraphics[scale=1.0]{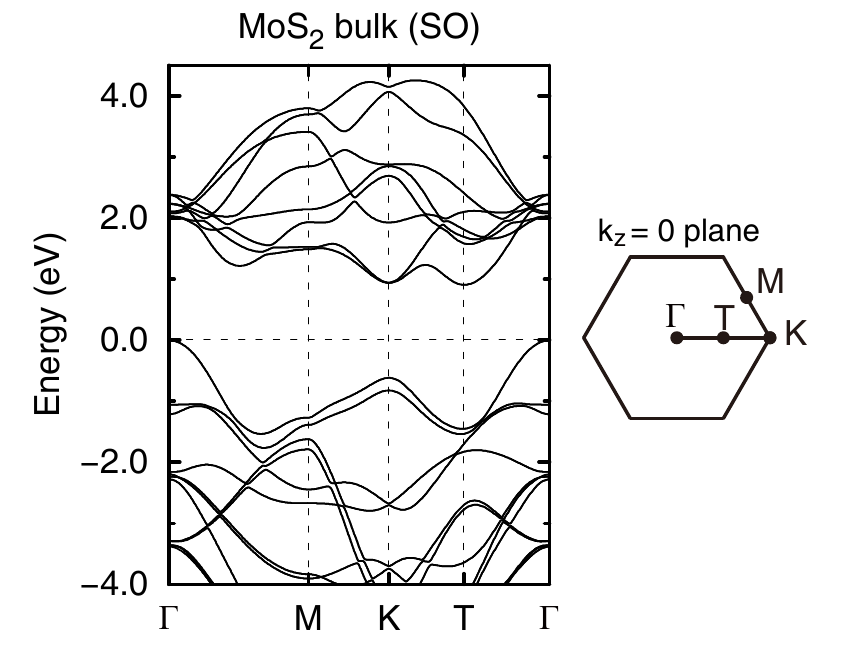}
\centering
\caption{\label{fig1tmdc} 
A prototypical band structure of TMD.
Electronic band structure of bulk MoS$_2$, which is obtained by DFT calculation including 
spin-orbit interaction, is plotted along high-symmetry lines.
VBM is set to zero energy.
The $k_z = 0$ plane of the three-dimensional (3D) BZ is also shown.
Location of the $T$ point, which is a local minimum of the lowest conduction band, depends on the
thickness and chemical compound of TMD, as shown in Table~\ref{tab3tmdc}.
}
\end{figure} 

We calculated electronic band structures of one-, two-, three-, four-layer, and bulk of 2H-phase $MX_2$ ($M =$ Mo, W and $X =$ S, Se, Te). 
Since all considered compounds have qualitatively similar band structures except for details,
we show the band structure of bulk MoS$_2$ in Fig.~\ref{fig1tmdc} as a prototype of thick samples. 
In Fig.~\ref{fig1tmdc}, the CBM of bulk MoS$_2$ is at a $k$-point in the $\Gamma$-$K$ line, which we define as the $T$ point, and the VBM is at the $\Gamma$ point. 
The location of the $T$ point in the $\Gamma$-$K$ line depends on the thickness and chemical compounds of $MX_2$ as shown in Table~\ref{tab3tmdc}. 
As is already well known, locations of conduction- and valence-band edges in the $k$-space are system-dependent so that the CBM is at the $T$ or $K$ point while the VBM is at the $\Gamma$ or $K$ point. 
In our DFT results (Table~\ref{tab4tmdc}), the CBM jumps from the $T$ point to the $K$ point and the VBM jumps from the $\Gamma$ point to the $K$ point as the number of layers decreases in all considered compounds. 
A similar feature occurs in our GW results (Table~\ref{tab5tmdc}) except for MoTe$_2$ and WTe$_2$. 
In these two compounds, our GW results show that the CBM is always at the $T$ point and the VBM is always at the $K$ point so that their band gaps are indirect for any thickness.

The spin-orbit interaction has an important role in electronic structures of 2H-phase TMDs. 
Even in the case of a compound of light elements such as monolayer MoS$_2$, the spin splitting is about 0.15~eV at VBM at the $K$ point. 
For monolayer WTe$_2$, which contains much heavier elements, the spin splitting is about 0.5~eV at VBM at the $K$ point. 
In the case of bulk, the spin-orbit coupling has relatively small effect compared with 2D layers. 
Also, the spin splitting in the conduction band is much smaller than that in the valence band because the enhancement of the spin splitting in 2D TMDs is due to the orbital angular momentum generated by the broken mirror symmetry \cite{Oh2017}. 
Thus, for few-layer systems, the giant spin splitting raises the VBM at the $K$ point, resulting in a direct band gap if the CBM is also at the $K$ point.

\begin{table} 
\caption[Location of the $T$ point in the $\Gamma$-$K$ line.]{\label{tab3tmdc} 
Location of the $T$ point in the $\Gamma$-$K$ line. 
The $T$ point is ${\bf k} = (t/3) {\bf b}_1 + (t/3) {\bf b}_2$ while the $K$ point is ${\bf k} = (1/3) {\bf b}_1 + (1/3) {\bf b}_2$. Here, ${\bf b}_1$ and ${\bf b}_2$ are in-plane reciprocal lattice vectors.
}
\setlength{\tabcolsep}{4.0mm} 
\renewcommand{\arraystretch}{1.12}
\begin{tabular}{l c c }

\hline
Compound \hspace{3mm} & Thickness  & \hspace{5mm} $t$ \hspace{5mm} \\
\hline
\multirow{2}*{MoS$_2$}   
    &   1L, 2L, 3L   &   0.52    \\
    &   4L, bulk   &   0.53    \\
\hline
\multirow{2}*{MoSe$_2$}   
    &   1L, 2L   &   0.54    \\
    &   3L, 4L, bulk   &   0.55    \\
\hline
\multirow{3}*{MoTe$_2$}   
    &   1L   &   0.55    \\
    &   2L   &   0.58    \\
    &   3L, 4L, bulk   &   0.59    \\
\hline
\multirow{2}*{WS$_2$}   
    &   1L   &   0.51    \\
    &   2L, 3L, 4L, bulk   &   0.52    \\
\hline
\multirow{2}*{WSe$_2$}   
    &   1L, 2L, 3L, 4L   &   0.54    \\
    &   bulk   &   0.55    \\
\hline
\multirow{1}*{WTe$_2$}   
    &   1L, 2L, 3L, 4L, bulk   &   0.59    \\
\hline

\end{tabular}
\end{table} 

\begin{figure*} 
\includegraphics[scale=1.0]{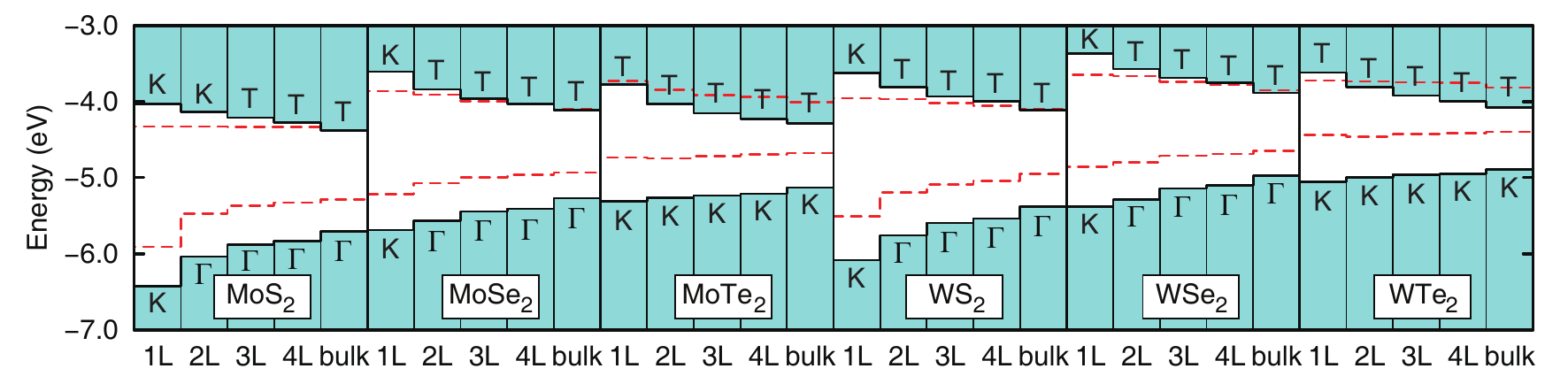}
\caption[Calculated band-edge energies of few-layer and bulk TMDs.]{\label{fig2tmdc} 
Calculated band-edge energies of few-layer and bulk TMDs. 
Band-edge energies are with respect to the vacuum level which is set to zero. 
Cyan regions show valence and conduction bands from our GW calculations, with characters $K$, $T$, and $\Gamma$ indicating locations of the CBM and VBM in the $k$-space. 
Red dashed lines show the CBM and VBM from our DFT calculations.}
\end{figure*} 

\begin{figure*} 
\includegraphics[scale=1.0]{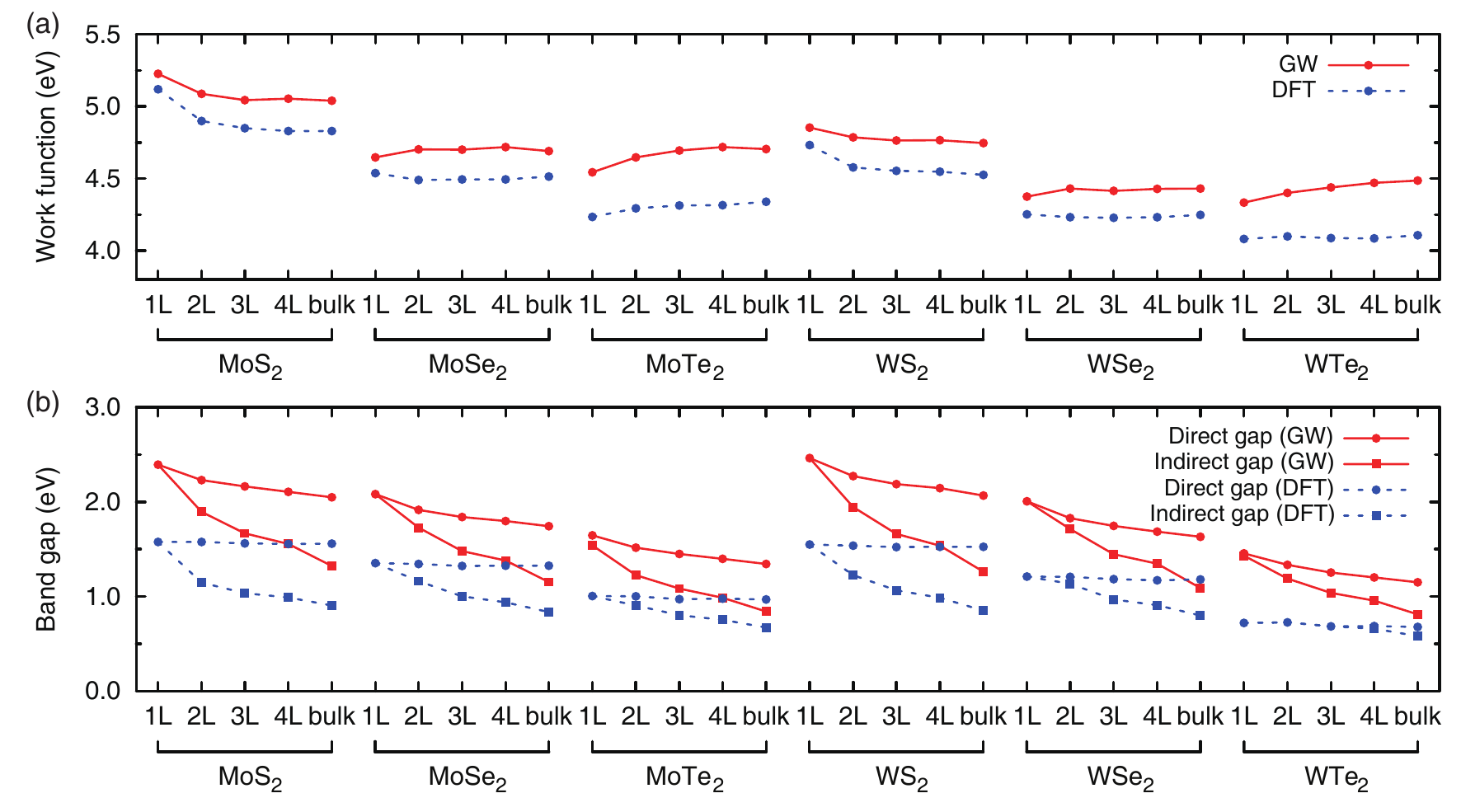}
\caption[Work functions of few-layer and bulk TMDs.]{\label{fig3tmdc} 
(a) Work functions of few-layer and bulk TMDs. 
Red (blue) dots are work functions obtained from our GW (DFT) calculations. 
(b) Band gaps of few-layer and bulk TMDs. 
Red (blue) circles and squares show direct and indirect band gaps, respectively, obtained from our GW (DFT) calculations.}
\end{figure*} 

\begin{table*} 
\caption{\label{tab4tmdc}
Band-edge energies, band gaps, work functions (WF), ionization potentials (IP), and electron affinities (EA) of TMDs from DFT calculations. 
As for the thickness, 1L, 2L, 3L, and 4L represent mono-, bi-, tri-, and tetralayer, respectively. 
Band-edge energies of $K_v$, $\Gamma$$_v$, $K_c$, and $T_c$ are for valence bands at $K$ and $\Gamma$ and conduction bands at $K$ and $T$, respectively, and they are with respect to the vacuum level which is set to zero. 
All values are in eV.}

\setlength{\tabcolsep}{3.7mm} 
\renewcommand{\arraystretch}{1.12}
\begin{tabular}{l c c c c c c c c c c}
\hline
\multirow{2}*{Compound}&\multirow{2}*{Thickness}&\multicolumn{4}{c}{Band energies}&\multicolumn{2}{c}{Band gaps}&\multirow{2}*{WF}&\multirow{2}*{IP}&\multirow{2}*{EA}\\
 & &K$_v$&$\Gamma$$_v$&K$_c$&T$_c$&Direct&Indirect& \\
\hline

\multirow{5}*{MoS$_2$}   &   1L   &   $-$5.91    &   $-$5.92    &   $-$4.33    &   $-$4.04    &   1.58    &       &   5.12    &   5.91    &   4.33    \\
   &   2L   &   $-$5.90    &   $-$5.47    &   $-$4.33    &   $-$4.20    &   1.58    &   1.15    &   4.90    &   5.47    &   4.33    \\
   &   3L   &   $-$5.89    &   $-$5.37    &   $-$4.33    &   $-$4.27    &   1.56    &   1.04    &   4.85    &   5.37    &   4.33    \\
   &   4L   &   $-$5.89    &   $-$5.32    &   $-$4.33    &   $-$4.30    &   1.56    &   0.99    &   4.83    &   5.32    &   4.33    \\
   &   bulk   &   $-$5.90    &   $-$5.28    &   $-$4.34    &   $-$4.38    &   1.56    &   0.91    &   4.83    &   5.28    &   4.38    \\
\hline
\multirow{5}*{MoSe$_2$}   &   1L   &   $-$5.21    &   $-$5.54    &   $-$3.86    &   $-$3.70    &   1.35    &       &   4.54    &   5.21    &   3.86    \\
   &   2L   &   $-$5.21    &   $-$5.07    &   $-$3.87    &   $-$3.91    &   1.34    &   1.16    &   4.49    &   5.07    &   3.91    \\
   &   3L   &   $-$5.20    &   $-$5.00    &   $-$3.88    &  $-$3.99    &   1.32    &   1.00    &   4.49    &   5.00    &   3.99    \\
   &   4L   &   $-$5.20    &   $-$4.96    &   $-$3.87    &   $-$4.03    &   1.33    &   0.94    &   4.49    &   4.96    &   4.03    \\
   &   bulk   &   $-$5.20    &   $-$4.93    &   $-$3.88    &   $-$4.10    &   1.32    &   0.84    &   4.51    &   4.93    &   4.10    \\
\hline
\multirow{5}*{MoTe$_2$}   &   1L   &   $-$4.74    &   $-$5.32    &   $-$3.73    &   $-$3.65    &   1.00    &       &   4.23    &   4.74    &   3.73    \\
   &   2L   &   $-$4.75    &   $-$4.77    &   $-$3.75    &   $-$3.84    &   1.00    &   0.90    &   4.29    &   4.75    &   3.84    \\
   &   3L   &   $-$4.72    &   $-$4.71    &   $-$3.75    &   $-$3.91    &   0.97    &   0.80    &   4.31    &   4.71    &   3.91    \\
   &   4L   &   $-$4.72    &   $-$4.69    &   $-$3.74    &   $-$3.94    &   0.98    &   0.75    &   4.31    &   4.69    &   3.94    \\
   &   bulk   &   $-$4.71    &   $-$4.68    &   $-$3.74    &   $-$4.00    &   0.97    &   0.67    &   4.34    &   4.68    &   4.00    \\
\hline
\multirow{5}*{WS$_2$}   &   1L   &   $-$5.51    &   $-$5.71    &   $-$3.96    &   $-$3.84    &   1.55    &       &   4.73    &   5.51    &   3.96    \\
   &   2L   &   $-$5.50    &   $-$5.19    &   $-$3.96    &   $-$3.94    &   1.54    &   1.23    &   4.58    &   5.19    &   3.96    \\
   &   3L   &   $-$5.50    &   $-$5.09    &   $-$3.98    &   $-$4.02    &   1.52    &   1.07    &   4.55    &   5.09    &   4.02    \\
   &   4L   &   $-$5.51    &   $-$5.04    &   $-$3.98    &   $-$4.05    &   1.52    &   0.99    &   4.55    &   5.04    &   4.05    \\
   &   bulk   &   $-$5.49    &   $-$4.95    &   $-$3.96    &   $-$4.10    &   1.53    &   0.85    &   4.53    &   4.95    &   4.10    \\
\hline
\multirow{5}*{WSe$_2$}   &   1L   &   $-$4.86    &   $-$5.29    &   $-$3.65    &   $-$3.53    &   1.21    &       &   4.25    &   4.86    &   3.65    \\
   &   2L   &   $-$4.87    &   $-$4.80    &   $-$3.66    &  $-$3.66    &   1.21    &   1.14    &   4.23    &   4.80    &   3.66    \\
   &   3L   &   $-$4.86    &   $-$4.71    &   $-$3.68    &   $-$3.74    &   1.18    &   0.97    &   4.23    &   4.71    &   3.74    \\
   &   4L   &   $-$4.85    &   $-$4.69    &   $-$3.68    &   $-$3.78    &   1.17    &   0.91    &   4.23    &   4.69    &   3.78    \\
   &   bulk   &   $-$4.86    &   $-$4.65    &   $-$3.69    &   $-$3.85    &   1.18    &   0.80    &   4.25    &   4.65    &   3.85    \\
\hline
\multirow{5}*{WTe$_2$}   &   1L   &   $-$4.44    &   $-$5.03    &   $-$3.72    &   $-$3.52    &   0.72    &       &   4.08    &   4.44    &   3.72    \\
   &   2L   &   $-$4.46    &   $-$4.50    &   $-$3.74    &   $-$3.64    &   0.73    &       &   4.10    &   4.46    &   3.74    \\
   &   3L   &   $-$4.43    &   $-$4.43    &   $-$3.75    &   $-$3.71    &   0.68    &       &   4.09    &   4.43    &   3.75    \\
   &   4L   &   $-$4.44    &   $-$4.42    &   $-$3.75    &   $-$3.74    &   0.69    &   0.66    &   4.08    &   4.42    &   3.75    \\
   &   bulk   &   $-$4.45    &   $-$4.40    &   $-$3.77    &   $-$3.82    &   0.68    &   0.58    &   4.11    &   4.40    &   3.82    \\

\hline
\end{tabular}
\end{table*} 

\begin{table*} 
\caption[Quasiparticle energies, band gaps, work functions (WF), ionization potentials (IP), and electron affinities (EA) of TMDs from GW calculations. ]{\label{tab5tmdc}
Quasiparticle energies, band gaps, work functions (WF), ionization potentials (IP), and electron affinities (EA) of TMDs from GW calculations. 
As for the thickness, 1L, 2L, 3L, and 4L represent mono-, bi-, tri-, and tetralayer, respectively. 
Quasiparticle energies of $K_v$, $\Gamma$$_v$, $K_c$, and $T_c$ are for valence bands at $K$ and $\Gamma$ and conduction bands at $K$ and $T$, respectively, and they are with respect to the vacuum level which is set to zero. 
All values are in eV.}

\setlength{\tabcolsep}{3.7mm} 
\renewcommand{\arraystretch}{1.12}
\begin{tabular}{l c c c c c c c c c c}
\hline
\multirow{2}*{Compound}&\multirow{2}*{Thickness}&\multicolumn{4}{c}{Quasiparticle energies}&\multicolumn{2}{c}{Band gaps}&\multirow{2}*{WF}&\multirow{2}*{IP}&\multirow{2}*{EA}\\
& &K$_v$&$\Gamma$$_v$&K$_c$&T$_c$&Direct&Indirect& \\

\hline
\multirow{5}*{MoS$_2$}   &   1L   &   $-$6.42    &   $-$6.51    &   $-$4.03    &   $-$3.82    &   2.40    &       &   5.23    &   6.42    &   4.03    \\
   &   2L   &   $-$6.37    &   $-$6.03    &   $-$4.14    &   $-$4.10    &   2.23    &   1.90    &   5.09    &   6.03    &   4.14    \\
   &   3L   &   $-$6.34    &   $-$5.88    &   $-$4.18    &   $-$4.21    &   2.17    &   1.67    &   5.04    &   5.88    &   4.21    \\
   &   4L   &   $-$6.34    &   $-$5.83    &   $-$4.23    &   $-$4.28    &   2.11    &   1.56    &   5.05    &   5.83    &   4.28    \\
   &   bulk   &   $-$6.26    &   $-$5.70    &   $-$4.21    &   $-$4.38    &   2.05    &   1.32    &   5.04    &   5.70    &   4.38    \\
\hline
\multirow{5}*{MoSe$_2$}   &   1L   &   $-$5.69    &   $-$6.08    &   $-$3.61    &   $-$3.52    &   2.08    &       &   4.65    &   5.69    &   3.61    \\
   &   2L   &   $-$5.64    &   $-$5.57    &   $-$3.73    &   $-$3.84    &   1.91    &   1.73    &   4.70    &   5.57    &   3.84    \\
   &   3L   &   $-$5.62    &   $-$5.44    &   $-$3.78    &   $-$3.96    &   1.84    &   1.48    &   4.70    &   5.44    &   3.96    \\
   &   4L   &   $-$5.62    &   $-$5.41    &   $-$3.82    &   $-$4.03    &   1.80    &   1.38    &   4.72    &   5.41    &   4.03    \\
   &   bulk   &   $-$5.51    &   $-$5.27    &   $-$3.77    &   $-$4.11    &   1.74    &   1.15    &   4.69    &   5.27    &   4.11    \\
\hline
\multirow{5}*{MoTe$_2$}   &   1L   &   $-$5.31    &   $-$5.95    &   $-$3.67    &   $-$3.77    &   1.65    &   1.54    &   4.54    &   5.31    &   3.77    \\
   &   2L   &   $-$5.26    &   $-$5.56    &   $-$3.74    &   $-$4.03    &   1.52    &   1.23    &   4.65    &   5.26    &   4.03    \\
   &   3L   &   $-$5.24    &   $-$5.41    &   $-$3.79    &   $-$4.15    &   1.45    &   1.08    &   4.69    &   5.24    &   4.15    \\
   &   4L   &   $-$5.21    &   $-$5.39    &   $-$3.81    &   $-$4.23    &   1.40    &   0.99    &   4.72    &   5.21    &   4.23    \\
   &   bulk   &   $-$5.13    &   $-$5.25    &   $-$3.78    &   $-$4.28    &   1.34    &   0.84    &   4.71    &   5.13    &   4.28    \\
\hline
\multirow{5}*{WS$_2$}   &   1L   &   $-$6.09    &   $-$6.32    &   $-$3.62    &   $-$3.58    &   2.46    &       &   4.85    &   6.09    &   3.62    \\
   &   2L   &   $-$6.01    &   $-$5.76    &   $-$3.74    &   $-$3.81    &   2.27    &   1.95    &   4.78    &   5.76    &   3.81    \\
   &   3L   &  $-$5.99    &   $-$5.59    &   $-$3.80    &   $-$3.93    &   2.19    &   1.66    &   4.76    &   5.59    &   3.93    \\
   &   4L   &   $-$5.98    &   $-$5.54    &   $-$3.83    &   $-$4.00    &   2.15    &   1.54    &   4.77    &   5.54    &   4.00    \\
   &   bulk   &   $-$5.90    &   $-$5.38    &   $-$3.83    &   $-$4.11    &   2.07    &   1.26    &   4.75    &   5.38    &   4.11    \\
\hline
\multirow{5}*{WSe$_2$}   &   1L   &   $-$5.38    &   $-$5.84    &   $-$3.37    &   $-$3.32    &   2.01    &       &   4.38    &   5.38    &   3.37    \\
   &   2L   &   $-$5.33    &   $-$5.29    &   $-$3.50    &   $-$3.57    &   1.83    &   1.71    &   4.43    &   5.29    &   3.57    \\
   &   3L   &   $-$5.29    &   $-$5.14    &   $-$3.55    &   $-$3.69    &   1.75    &   1.45    &   4.41    &   5.14    &   3.69    \\
   &   4L   &   $-$5.28    &   $-$5.10    &   $-$3.59    &   $-$3.76    &   1.69    &   1.35    &   4.43    &   5.10    &   3.76    \\
   &   bulk   &   $-$5.23    &   $-$4.98    &   $-$3.60    &   $-$3.89    &   1.63    &   1.09    &   4.43    &   4.98    &   3.89    \\
\hline
\multirow{5}*{WTe$_2$}   &   1L   &   $-$5.05    &   $-$5.66    &   $-$3.59    &   $-$3.62    &   1.45    &   1.43    &   4.33    &   5.05    &   3.62    \\
   &   2L   &   $-$5.00    &   $-$5.24    &   $-$3.66    &   $-$3.81    &   1.33    &   1.19    &   4.40    &   5.00    &   3.81    \\
   &   3L   &   $-$4.96    &   $-$5.07    &   $-$3.71    &   $-$3.92    &   1.25    &   1.04    &   4.44    &   4.96    &   3.92    \\
   &   4L   &   $-$4.95    &   $-$5.06    &   $-$3.75    &   $-$3.99    &   1.20    &   0.96    &   4.47    &   4.95    &   3.99    \\
   &   bulk   &   $-$4.89    &   $-$4.92    &   $-$3.74    &   $-$4.08    &   1.15    &   0.81    &   4.49    &   4.89    &   4.08    \\
\hline
\end{tabular}
\end{table*} 

\begin{table} 
\caption{\label{tab6tmdc} 
Direct ($E_g^{(d)}$) and indirect ($E_g^{(i)}$) band gaps (in eV) 
of bulk TMDs obtained from our PBE and GW calculations.}
\setlength{\tabcolsep}{0.5mm} 
\renewcommand{\arraystretch}{1.12}
\begin{tabular}{c c c c c l}
\hline
Bulk & &\multicolumn{2}{c}{Our results}&\multirow{2}*{GW\cite{Jiang2012}}&\multirow{2}*{Experiment}\\
TMDs & &PBE&GW& & \\

\hline
\multirow{2}*{MoS$_2$} & $E_{g}^{(i)}$ & 0.91 & 1.32 & 1.23 & 1.22 \cite{Baglio1982}, 1.23 \cite{Kam1982}\\
 & $E_{g}^{(d)}$ & 1.56 & 2.05 & 2.07 & 1.77 \cite{Baglio1982}, 1.74 \cite{Kam1982} \\
\hline

\multirow{2}*{MoSe$_2$} & $E_{g}^{(i)}$ & 0.84 & 1.15 & 1.11 & 1.10 \cite{Baglio1982}, 1.12 \cite{Kam1984}, 1.09 \cite{Kam1982} \\
 & $E_{g}^{(d)}$ & 1.33 & 1.74 & 1.83 & 1.38 \cite{Baglio1982,Kam1982} \\
\hline

\multirow{2}*{MoTe$_2$} & $E_{g}^{(i)}$ & 0.67 & 0.84 &   & 0.88 \cite{Lezama2014}\\
 & $E_{g}^{(d)}$ & 0.97 & 1.34 &  & 1.02 \cite{Lezama2014} \\
\hline

\multirow{2}*{WS$_2$} & $E_{g}^{(i)}$ & 0.85 & 1.26 & 1.30 & 1.34 \cite{Baglio1982}, 1.35 \cite{Kam1982}\\
     & $E_{g}^{(d)}$ & 1.53 & 2.07 & 2.13 & 1.78 \cite{Baglio1982}, 1.79 \cite{Kam1982}, 2.01 \cite{ballif} \\
\hline

\multirow{2}*{WSe$_2$} & $E_{g}^{(i)}$ & 0.80 & 1.09 & 1.19 & 1.22 \cite{Kam1984}, 1.20 \cite{Kam1982}, 1.2 \cite{Traving1997} \\
 & $E_{g}^{(d)}$ & 1.18 & 1.63 & 1.75 & 1.39 \cite{Kam1982}, 1.75 \cite{Beal1976} \\
\hline

\multirow{2}*{WTe$_2$} & $E_{g}^{(i)}$ & 0.58 & 0.81 &  & \\
 & $E_{g}^{(d)}$ & 0.68 & 1.15 &  &  \\
\hline
\end{tabular}
\end{table} 

Band gaps of bulk TMDs from our PBE and GW calculations are shown in Tables~\ref{tab4tmdc} and \ref{tab5tmdc}, respectively, and compared with previous calculational and experimental results in Table~\ref{tab6tmdc}. 
In these tables, we define the {\em direct} band gap as the minimal energy difference between valence bands and conduction bands at the same k point, while the band gap is defined conventionally as the energy difference between the CBM and the VBM. 
If the CBM and the VBM are located at different $k$ points, the band gap will be termed as the {\em indirect} band gap in our present work. 
We confirm the accuracy of our study by comparing our results of bulk band gaps with previous reports \cite{Jiang2012,Baglio1982,Kam1984,Kam1982,Beal1976,Traving1997,Lezama2014,ballif}. As shown in Table~\ref{tab6tmdc}, our results of bulk band gaps from the GW method are in good agreement with previous theoretical results \cite{Jiang2012} within 0.12~eV. When compared with experiments, our GW results of indirect band gaps agree with experimental ones \cite{Baglio1982,Kam1984,Kam1982,Traving1997,Lezama2014} within 0.13~eV, while values of direct band gaps show a little bit more deviation from experimental ones \cite{Baglio1982,Kam1982,Beal1976,Lezama2014,ballif}. 
More detailed comparison shows that our GW results of direct band gaps are in better agreement with experimental results considering excitonic effects \cite{Beal1976,ballif}.

The CBM and the VBM of few-layer and bulk TMDs from our DFT and GW calculations are shown in Tables~\ref{tab4tmdc} and \ref{tab5tmdc}, respectively, and plotted in Fig.~\ref{fig2tmdc}. 
Comparing GW results with DFT results, we find that the CBM depends more strongly on the thickness in the GW results than in the DFT results. 
In the case of VBM, the self-energy correction from the GW method is almost constant near the VBM, making the difference between DFT and GW results almost independent of the thickness. 
In our GW results, the thickness dependence of the CBM is almost independent of the chemical composition of TMDs while the thickness dependence of the VBM becomes weaker in the order of sulfides, selenides, and tellurides.

Work functions of few-layer and bulk TMDs obtained from our DFT and GW calculations are shown in Tables~\ref{tab4tmdc} and \ref{tab5tmdc}, respectively, and plotted in Fig.~\ref{fig3tmdc}(a). 
As shown in Fig.~\ref{fig3tmdc}(a), work functions from GW calculations are larger than those from DFT calculations.
This enhancement of the work function is more pronounced in compounds containing Te elements. 
The thickness dependence of the work function is very weak except for the monolayer. 
In our GW results, the work function is the largest for MoS$_2$, followed by WS$_2$, MoSe$_2$ and MoTe$_2$, and WSe$_2$ and WTe$_2$. 
Thus, sulfides have larger work functions than selenides and tellenides. 
With the same chalcogen element, Mo compounds have larger work functions than W compounds. 
We also obtain ionization energies and electron affinities from band-edge energies with respect to the vacuum level in DFT and GW results, as shown in Tables~\ref{tab4tmdc} and \ref{tab5tmdc}, respectively.

Band gaps of few-layer TMDs obtained from our DFT and GW calculations are shown in Tables~\ref{tab4tmdc} and \ref{tab5tmdc}, respectively, and plotted in Fig.~\ref{fig3tmdc}(b). 
Overall, we note that band gaps decrease in the order of sulfides, selenides, and tellurides. 
In our GW results, band gaps are weakly dependent on metal elements but strongly dependent on chalcogen elements. 
Few-layer and bulk MoS$_2$ have almost the same band gaps as WS$_2$. Band gaps in MoSe$_2$ (MoTe$_2$) are close to those in WSe$_2$ (WTe$_2$) within 0.2~eV. Thus we conclude that band gaps of TMDs in 2H phase are almost determined by the chalcogen element.

In our GW results, shown in Fig.~\ref{fig3tmdc}(b), direct band gaps have significant thickness dependence, while they hardly depend on the thickness in our DFT results. 
As for indirect band gaps, our GW results show larger thickness dependence than our DFT results. 
In our GW results, the band gap is direct in monolayer MoS$_2$, MoSe$_2$, WS$_2$, and WSe$_2$ only, and it is indirect in all other cases. 
In our DFT results, the band gap is direct in all monolayers.

\begin{figure} [b]  
\includegraphics[scale=1.0]{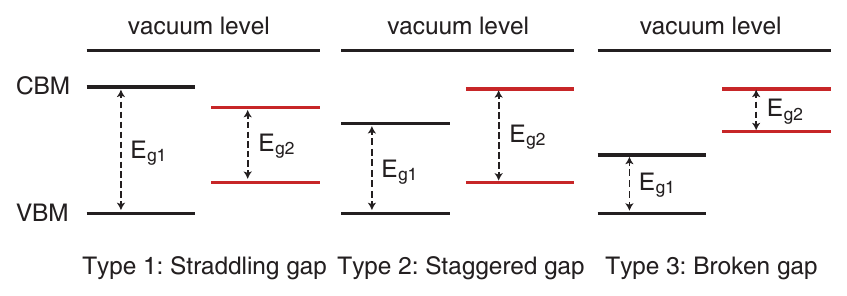}
\centering
\caption{\label{fig4tmdc} 
Types of band alignments in junctions of two different semiconductors.}
\centering
\end{figure} 

Now we discuss band alignments in junctions of different TMDs. 
In general, as shown in Fig.~\ref{fig4tmdc}, there are three types of band alignments: straddling, staggered, and broken band gaps, which we denote by type 1, type 2, and type 3, respectively. 
The straddling type (type 1) is the case that the band gap of one material is placed entirely inside the band gap of the other material in energy. 
The staggered type (type 2) is the case that band gaps of two materials overlap partially in energy. 
The broken type (type 3) is the case that band gaps of two materials do not overlap in energy at all.
We consider 375 junctions of two different compounds, where each compound is one-, two-, three-, four-layer, or bulk. 
We obtain the number of possible types of band alignments, as shown in Table~\ref{tab7tmdc},
by estimating types of band alignments of junctions 
from our DFT band energies (Table~\ref{tab4tmdc})
and GW quasiparticle energies (Table~\ref{tab5tmdc}) of each compound,
assuming that band-edge energies of a constituent material with respect to the vacuum level are not changed by the presence of the other constituent material. 
This approach is more relevant for in-plane junctions where two compounds are connected side-by-side, while it may have less validity for out-of-plane junctions where one compound is stacked on top of the other. 
From our DFT band energies of each compound given in Table~\ref{tab4tmdc}, 65 junctions have the straddling type, 310 junctions have the staggered type, and none have the broken type. 
When the band-edge energies are corrected using the GW method, 92 junctions are changed from the staggered type to the straddling type and 7 junctions are changed from the straddling type to the staggered type. 
Thus, from our GW results of each compound given in Table~\ref{tab5tmdc}, 150 junctions have the straddling type, 225 junctions have the staggered type, and none has the broken type.
In the case of out-of-plane junctions, each constituent compound can contribute to screening of holes and electrons in the other compound, and this raises GW valence bands and lowers GW conduction bands \cite{neaton} in the other compound. Since this raising and lowering occurs in both compounds of a junction, the band alignment will be affected only by the difference in their screening effects, and the difference will be small when the two compounds have similar dielectric response.

\begin{figure} 
\includegraphics[scale=1.0]{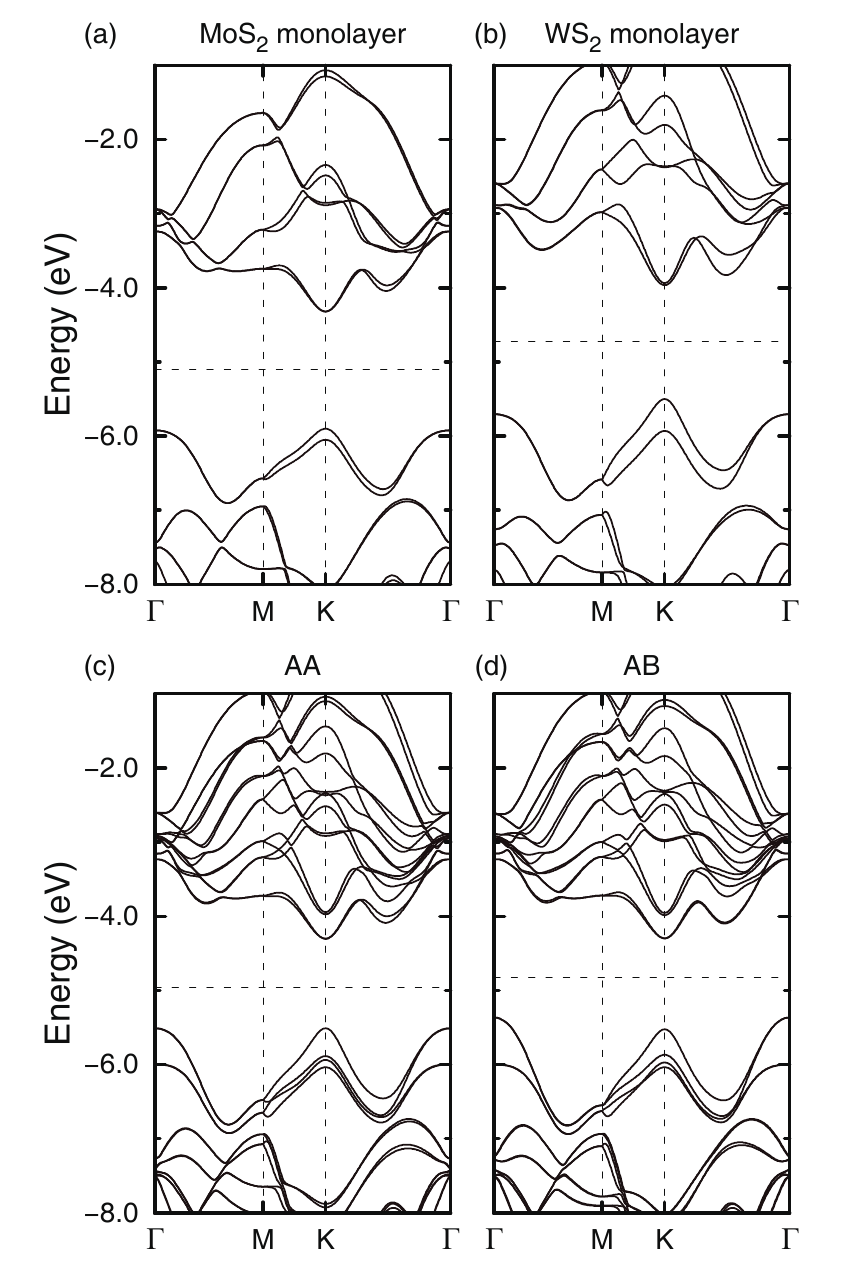}
\centering
\caption[DFT band structures of monolayer MoS$_2$, monolayer WS$_2$, AA-stacked out-of-plane MoS$_2$/WS$_2$ junction, and AB-stacked out-of-plane MoS$_2$/WS$_2$ junction. ]{\label{fig5tmdc} 
DFT band structures of (a) monolayer MoS$_2$, (b) monolayer WS$_2$, (c) AA-stacked out-of-plane MoS$_2$/WS$_2$ junction, and (d) AB-stacked out-of-plane MoS$_2$/WS$_2$ junction. 
AA- and AB-stacked out-of-plane MoS$_2$/WS$_2$ junctions are made by placing monolayer MoS$_2$ on top of monolayer WS$_2$ with AA and AB stacking, respectively. 
Band energies are with respect to the vacuum level which is set to zero. Horizontal dashed lines indicate the center of the band gaps.
}
\end{figure}

To test an interlayer interaction effect in out-of-plane junctions, we consider out-of-plane junctions of MoS$_2$ and WS$_2$ by placing monolayer MoS$_2$ on top of monolayer WS$_2$ with two types of stacking, AA and AB. 
Fig.~\ref{fig5tmdc} shows band structures of monolayer MoS$_2$, monolayer WS$_2$, and their AA- and AB-stacked out-of-plane junctions, respectively, obtained by DFT calculations. 
By comparing band structures of the two junctions [Figs.~\ref{fig5tmdc}(c) and \ref{fig5tmdc}(d)] with those of monolayers [Figs.~\ref{fig5tmdc}(a) and \ref{fig5tmdc}(b)], we find that band structures of out-of-plane junctions are very similar to the simple superposition of band structures of monolayers except for valence bands near the $\Gamma$ point. 
In the AB-stacked junction, valence bands at the $\Gamma$ point are split greatly due to interlayer interaction, shifting the VBM from the $K$ point to the $\Gamma$ point [Fig.~\ref{fig5tmdc}(d)]. 
Meanwhile, in the AA-stacked junction, splitting of valence bands at the $\Gamma$ point is rather weak, remaining VBM at the $K$ point [Fig.~\ref{fig5tmdc}(c)]. 
Thus, interlayer interaction may affect band alignments in out-of-plane TMD junctions by changing valence-band dispersion near the $\Gamma$ point.

\begin{table} 
\caption{\label{tab7tmdc} 
Types of band alignments in TMD junctions.
The number of TMD junctions is shown for each type of band alignment.
The band alignment in each junction is determined by comparing VBMs and CBMs of constituent parts 
which are obtained with respect to the vacuum level by DFT and GW calculations.}
\setlength{\tabcolsep}{1.3mm} 
\renewcommand{\arraystretch}{1.12}
\begin{tabular}{c c c c c}
\hline
 &Straddling type&Staggered type&Broken type&Total \\
\hline
DFT&65&310&0&375\\
 GW&150&225&0&375 \\
\hline
\end{tabular}
\end{table} 

In these out-of-plane MoS$_2$/WS$_2$ junctions we matched the in-plane lattice constants of MoS$_2$ and WS$_2$ by expanding the in-plane lattice constant of MoS$_2$ by 0.1\%. 
In general, lattice-constant mismatch, any rotation or twisting \cite{Heo2015}, and/or intercalation of insulating layers such as h-BN \cite{Pant2016} between constituent TMDs in out-of-plane junctions may change the strength of interlayer interaction, and thereby affect band alignments in TMD junctions. 
Our main results, which are band-edge energies of few-layer and bulk TMDs with respect to the vacuum level obtained from GW calculations, can be used to predict band alignments of out-of-plane TMD junctions when interlayer interaction is weak.

\section{Conclusion}

We studied electronic band structures of few-layer and bulk TMDs in 2H phase (MoS$_2$, MoSe$_2$, MoTe$_2$, WS$_2$, WSe$_2$, and WTe$_2$) using DFT and GW calculations. 
We considered one-, two-, three-, four-layer, and bulk geometry of each compound. 
We calculated VBMs and CBMs with respect to the vacuum level by DFT and GW calculations, obtaining band gaps, work functions, ionization energies, and electron affinities from VBMs and CBMs for few-layer and bulk geometry. 
Obtained CBM with respect to the vacuum level depends more strongly on the thickness in GW results than in DFT results. 
In the case of VBM with respect to the vacuum level, the difference between DFT and GW results is found almost independent of the thickness. 
Work functions from GW calculations are larger than those from DFT calculations, and their thickness dependence is very weak except for monolayer.
Band gaps from GW calculations have stronger thickness dependence than those from DFT calculations.
Our GW calculations show that band gaps are direct in monolayer MoS$_2$, MoSe$_2$, WS$_2$, and WSe$_2$ only, and they are indirect in all other cases. 
Based on the electronic structure of each TMD, we discussed types of band alignments in in-plane and out-of-plane junctions of different TMDs,
finding some difference in band-alignment types when we compared DFT and GW results. 
We also examined out-of-plane MoS$_2$/WS$_2$ junctions with different stacking, showing that interlayer interaction, unless weakened by some exotic geometry, may affect band alignments in out-of-plane TMD junctions by changing the valence-band dispersion near the $\Gamma$ point.

\begin{acknowledgments}
This work is supported by the National Research Foundation of Korea (Grant No. 2020R1A2C3013673). 
Computational resources have been provided by KISTI Supercomputing Center (Project No. KSC-2019-CRE-0195).
\end{acknowledgments}

\end{document}